\documentclass[aps,prl,twocolumn,showpacs,superscriptaddress,amsmath,amssymb,floatfix,altaffillsymbol]{revtex4-1}
\usepackage{graphicx}
\usepackage{ulem}
\usepackage{dcolumn}
\usepackage{bm}
\usepackage{amsmath}
\usepackage{latexsym}
\usepackage{amsfonts}
\usepackage{hyperref}
\usepackage{amssymb}
\usepackage{verbatim}
\usepackage[rightcaption]{sidecap}
\usepackage{color}
\usepackage{soul}
\usepackage{verbatim}
\usepackage[export]{adjustbox}
\usepackage{tikz}
\usetikzlibrary{shapes.geometric, arrows}
\usetikzlibrary{decorations.markings}
\usetikzlibrary{decorations.pathmorphing}
\usepackage{mathtools, nccmath, relsize}
\usepackage[percent]{overpic}

\begin{document}

\title{Density Functional Theory of the Hubbard-Holstein Model}
\author{E. Vi\~nas Bostr\"om}
\affiliation{Department of Physics, Lund University, PO Box 118, 221 00 Lund, Sweden}
\author{P. Helmer}
\affiliation{Department of Physics, Lund University, PO Box 118, 221 00 Lund, Sweden}
\author{P. Werner}
\affiliation{Department of Physics, University of Fribourg, 1700 Fribourg, Switzerland}
\author{C. Verdozzi}
\affiliation{Department of Physics, Lund University, PO Box 118, 221 00 Lund, Sweden}
\date{\today}

\begin{abstract}
We present a density functional theory (DFT) for lattice models with local electron-electron (e-e) and electron-phonon (e-ph) interactions. Exchange-correlation potentials are derived via dynamical mean field theory for the infinite-dimensional Bethe lattice, and analytically for an isolated Hubbard-Holstein site. These potentials exhibit discontinuities as a function of the density, which depend on the relative strength of the e-e and e-ph interactions. By comparing to exact benchmarks, we show that the DFT formalism gives a good description of the linear conductance and real-time dynamics.
\end{abstract}
\maketitle


Density functional theory (DFT)~\cite{HK,KS65} plays a central role in the study of materials~\cite{Review}. The key idea of DFT and its time-dependent generalization (TDDFT)~\cite{RG84} is to focus on the one-particle density as the basic variable. Although DFT and TDDFT are in principle exact methods, in practice the exchange-correlation (XC) potential, a key ingredient in the theory, is in most cases known only approximately. (TD)DFT has also been applied to model Hamiltonians~\cite{GunSchon84,KCPillars}, to explore conceptual and methodological aspects of the theory~\cite{Lima,Arya1,CV08,Baer,Ullrich,GSPECI,Concept1,Maitra,Concept2,Concept3,sanvito,franca}, 
but also for specific applications to e.g. cold atoms~\cite{KCcold1,KCcold2,DKcold,AKcold}, Kondo physics~\cite{GSSK,Burke,Evers}, quantum transport~\cite{Kurth10,SK1,DK}, quantum electrodynamics~\cite{MFIT}, and nonequilibrium thermodynamics \cite{Damico}, to mention a few. There is however a class of lattice systems not considered until now within (TD)DFT: electron-phonon lattice models. This is surprising, given that the interplay of electron-electron (e-e) and electron-phonon (e-ph) interactions is at the root of many physical phenomena. (TD)DFT could provide new insights into the correlation effects in these systems, and in particular the effect of phonons on the XC potentials.

Here, we introduce a (TD)DFT description of a paradigmatic e-ph lattice system, the Hubbard-Holstein (HH) model~\cite{Freericks1995, SupercondCDW, MIT, Hewson, Sangiovanni2005, Bauer2010, HHrefs1,HHrefs2,HHrefs3, HHrefs4,HHrefs5}. This model provides a minimum-complexity set-up to treat on equal footing e-e and e-ph interactions in a lattice.  Still, depending on model parameters and dimensionality $D$, it displays a broad range of interesting behaviors, e.g. metal-insulator transitions~\cite{MIT}, bipolaronic phases~\cite{Hewson}, superconductivity and charge density order~\cite{SupercondCDW}. In the out of equilibrium regime, it has been used to study interaction quenches~\cite{quenches}, dynamical insulator-to-metal transitions~\cite{Strand}, and pump-probe dynamics~\cite{DeFilippis,Werner2015}. Finally, through phonon overscreening of the e-e interactions, it provides a physical motivation to study lattice models with attractive Hubbard interactions~\cite{NegUHubb,Campo,Perfetto12}.
 
Similarly to what is done in (TD)DFT for electron-nuclei systems in the continuum~\cite{Gross_eph,RvL_ph2}, or quantum electrodynamics~\cite{MFIT}, we describe the HH model via a two-component formulation for the electron and phonon subsystems, where for the electron (phonon) component the basic variables are the electron occupations $\{n_i\}$ 
(phonon coordinates $\{x_i\}$) at each site $i$, and each component is governed by its own XC potential.

After presenting our approach, we explicitly determine the XC potentials for the analytically solvable one-site, zero-dimensional ($D=0$) model, and the infinite dimensional ($D=\infty$) homogenous Bethe lattice (where $n_i=n, x_i=x$), via dynamical mean field theory (DMFT)~\cite{DMFTrefs1,DMFTrefs2,PhilippDMFT}. These potentials are then used
to compute the dynamics in a finite system. We find that: i) e-ph interactions screen the e-e interaction, and the behavior of the electronic XC potential is mainly determined by the screened interaction $U'$. At the same time,  because the e-ph coupling is linear in both $n$ and $x$, as well as local, the XC phonon potential is always zero. ii) The electronic XC potential is discontinuous at half-filling density $n=1$ for $U'>0$, and at $n=0,2$ for $U'<0$;  for the Bethe lattice, the discontinuity appears above a 
nonzero
value of $|U'|$.
iii) For an infinite chain with a HH impurity, (TD)DFT conductances have a smooth transition from the charge- to the spin-Kondo regime upon varying the e-ph coupling. iv) TDDFT dynamics in a test system subject to interaction quenches or external fields compares well with exact numerics in an appreciable range of interaction strengths. These results demonstrate that (TD)DFT  is a promising formalism for the study of e-ph lattice systems.


{\it The system -} 
Aiming for a DFT description, we start with an inhomogeneous version of the HH Hamiltonian:
\begin{eqnarray}\label{eq:ham}
&&\!\!\!\!\!\hat{H} =\sum_{i\sigma}  (v_i-\mu) \hat{n}_{i\sigma} + U\sum_i \hat{n}_{i\uparrow} \hat{n}_{i\downarrow} - J\sum_{\langle ij\rangle \sigma}  c_{i\sigma}^\dagger c_{j\sigma}  \nonumber \\
 &&\!\!\!\!\!+\;\omega\sum_i b_i^\dagger b_i +\sum_i \sqrt{2}\eta_i\;\hat{x}_i + \sqrt{2}g
\sum_i(\hat{n}_{i\uparrow}\!+\!\hat{n}_{i\downarrow}\!-\!1)
 \hat{x}_i,~~~
\end{eqnarray}
where $v_i$ is a local site dependent electron potential, $\mu$ is the chemical potential, $U$ the e-e interaction strength, $J$ the hopping amplitude (set equal to 1, as the energy unit), and $\langle \dots\rangle$ denotes nearest neighbour sites. The operator $c^\dagger_{i\sigma}$  creates an electron on site $i$ with spin $\sigma$, with $\hat{n}_{i\sigma}=c_{i\sigma}^\dagger c_{i\sigma}$ the corresponding density operator. The phonon frequency is $\omega$, and $g$ is the e-ph coupling parameter. We use $\lambda = g^2/\omega$ as a measure of the e-ph interaction strength. Finally, the site-dependent external phonon potential $\eta_i$ is introduced to control the phonon coordinates $\hat{x}_i = (b_i^\dagger + b_i)/\sqrt{2}$, where $b_i$ destroys a phonon at site $i$.  The form of $\hat{H}$ in Eq.~(\ref{eq:ham}) 
allows 
to address formal aspects of the (TD)DFT description (see the Supplemental Material, SM), and 
to use a homogeneous HH
reference system ($\forall i$, $v_i=v$ and $\eta_i=\eta$) in (adiabatic) local density approximations. To calculate the ground state energy of the reference homogenous HH model, we perform the Lang-Firsov transformation $H\rightarrow \hat{H}'= e^{i\hat{S}}\hat{H}e^{-i\hat{S}}$ \cite{LangFirs} (see the SM). In $H'$, the hopping amplitude is renormalized as $J\rightarrow \hat{J}'_{ij} = Je^{i\sqrt{2}g(\hat{p}_i-\hat{p}_j)/\omega}$, 
with $\hat p_i = i(b_i^\dagger - b_i)/\sqrt{2}$ the phonon momentum,
and the other parameters transform as $v \rightarrow v' = v + (g^2 + 2g\eta)/\omega$, and $U \rightarrow U' = U - 2g^2/\omega$.


{\it Density functional theory -} 
For a DFT description, we consider the pair of sets of variables $(n,x) \equiv (\{n_i\},\{x_i\})$ and the conjugated   fields $(v,\eta) \equiv (\{v_i\},\{\eta_i\})$, with $n_i=n_{i\uparrow}+n_{i\downarrow}$ the total electron density at site $i$. In the SM we prove that i) the total energy $E = E[n,x]$ is a functional of $n$ and $x$, with a minimum at the ground state values $(n_0,x_0)$ (i.e., the Hohenberg-Kohn theorem for the HH) and ii) where $v_0$ representability holds, $(n_0,x_0)$ is obtained by solving a two-component Kohn-Sham (KS) problem. In a homogeneous system (useful to derive a local density approximation), $n_i=n$ and $x_i=x$ for all $i$, and the KS Hamiltonian is $H_{KS}=H_s^{(e)}+H_s^{(ph)}$, with
\begin{align}
&H_s^{(e)} = (v_{KS}[n,x]-\mu)\sum_{i\sigma} \hat{n}_{i\sigma} -  \sum_{\langle ij\rangle \sigma}J\left(c_{i\sigma}^\dagger c_{j\sigma} + h.c. \right), \nonumber \\
&H_s^{(ph)}= \omega\sum_i b_i^\dagger b_i + \sqrt{2}\eta_{KS}[n,x]\sum_i \hat{x}_i~.
\end{align}
The  electronic KS potential can be written as $v_{KS} = v_{ext} + v_{Hxc}$, with $v_{ext} \equiv v$ and the Hartree-exchange-correlation (Hxc) part $v_{Hxc} = Un/2 + \sqrt{2}gx + \delta E_{xc}/\delta n$. To obtain $E_{xc}$, we subtract  from  $E[n,x]$ the e-e and e-ph Hartree interaction terms, and the energies of the corresponding non-interacting HH system. For the phonons, $\eta_{KS}= \eta_{ext}+ \eta_{Hxc}$, where $\eta_{ext} \equiv \eta$ and $\eta_{Hxc} \equiv \eta_H+\eta_{xc} = \sqrt{2}g(n-1) + \delta E_{xc}/\delta x$. Using the Heisenberg equation $\partial_t\hat{p} = 0$, we get $x = -\sqrt{2}\left[g(n-1) + \eta\right]/\omega$. Inverting this relation and using $\eta_{Hxc} = \eta|_{g=0} - \eta$ , we find $\eta_{Hxc} = \eta_H$, i.e. $\eta_{xc} = 0$.
In the following, we set $\omega = 1$ and $\eta = 0$.


\begin{figure}
 \includegraphics[width=\columnwidth]{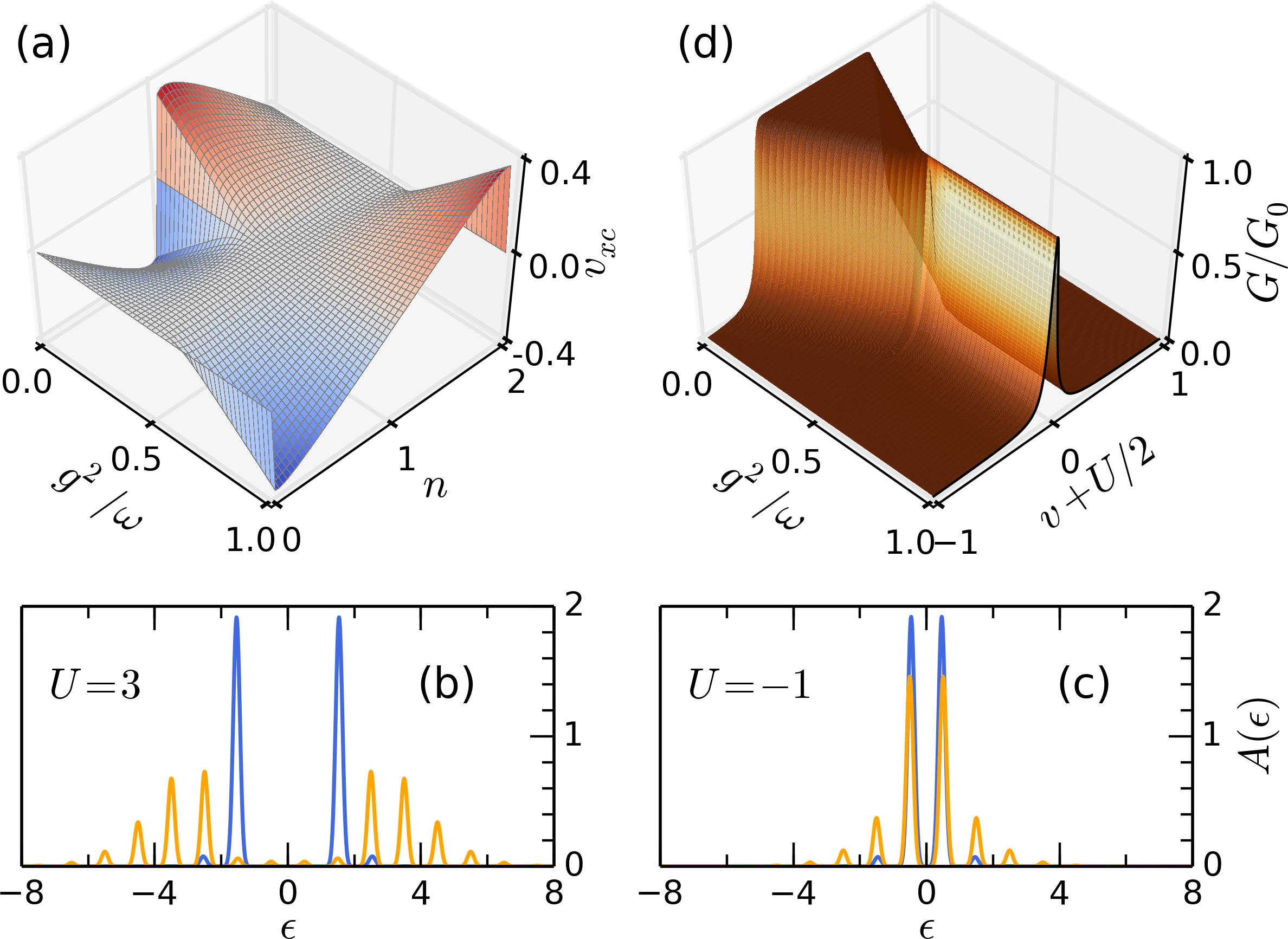}
 \caption{Color online. Properties of a single Hubbard-Holstein site: (a) Exchange-correlation potential $v_{xc}$ for $U = 1$ and $\beta = 20$, as a function of $n$ and $\lambda=g^2/\omega$. (b, c) Spectral function $A(\epsilon)$ for $\mu = 0$, $v = -U/2$ and $\beta = 5$. The blue curves are for $\lambda = 0.08$ and yellow curves for $\lambda = 1$. (d) Zero bias conductance $G$ for $U = 1$, $\mu = 0$ and $\beta \to \infty$, as a function of $v$ and $\lambda$.}
 \label{fig:single_site}
\end{figure}

{\it A) $v_{xc}$ from a single Hubbard-Holstein site -} 
We consider a single HH site exchanging energy and particles with a bath at chemical potential $\mu$ and temperature $\beta^{-1}$. This problem is analytically solvable, by determining the partition function $Z = {\rm Tr}\, e^{-\beta H(v,\eta)}$ (see SM). Using $n = -\beta^{-1}\partial_v \ln Z$, $x = -(\sqrt{2}\beta)^{-1}\partial_\eta\ln Z$, and solving for $(v,\eta) $ we find $\eta_{xc} = 0$ and
\begin{align}\label{eq:vxc}
v_{xc}(n,x) = (1-\delta n)\frac{U'}{2} + \frac{g^2}{\omega}\delta{n}+\frac{1}{\beta}\ln\frac{\delta n+ R}{1+\delta n},
\end{align}
where $\delta n = n - 1$, $R=[e^{-\beta U'}\left(1-\delta n^2\right)+\delta n^2]^{1/2}$, and $U'$ defined above. Thus,  $v_{xc}$ is independent of $x$ and depends on the e-e interaction renormalized by phonons. For $g=0$ we recover the expression for a single-site Hubbard system~\cite{GSSK}. In Fig.~\ref{fig:single_site}(a) we display $v_{xc}$ as a function of $n$ and $\lambda$ for $U = 1$ and $\beta = 20$. For $U' > 0$ (i.e. $\lambda < 1/2$) the potential is discontinuous at $n=1$,  as in the case of the purely electronic repulsive Hubbard model~\cite{GunSchonNoack,Lima,DKAPCV,GSSK}. For $U' < 0$ (i.e. $\lambda > 1/2$) the discontinuity is at $n = 0$ and $n = 2$, as in a negative-$U$ Hubbard model~\cite{KCcold2,Perfetto12}. Notably, in the present model, the transition from positive to negative $U'$ results from the phonons screening of the e-e interactions.
Eq.~(\ref{eq:vxc}) shows that, save for the linear term $g^2 \delta n/\omega$, the analytic expressions for $v_{xc}$ in a HH and a Hubbard single-site model only differ by the renormalization $U\rightarrow U'$, i.e e-ph interactions primarily affect the discontinuities at $n = \{0,1, 2 \}$. Phonon effects are instead explicitly manifest in the electronic spectral function $A(\epsilon)$. Starting from the many-body Matsubara Green's function $G^M(\tau) = -iZ^{-1}{\rm Tr}\,\left(e^{-\beta (\hat{H}-\mu \hat{N})}{\mathcal T}[c(\tau)c^\dagger(0)]\right)$, $A(\epsilon)$ can be extracted via analytic continuation to real energies and the fluctuation-dissipation theorem (see SM). 
Fig.~\ref{fig:single_site}(b,c) shows $A(\epsilon)$ for $\mu = 0$, $\beta = 5$, $v = -U/2$ and for four pairs $(U,\lambda)$. For $U = 3$ and $\lambda = 0.08$ the two main peaks correspond to the electronic excitation energies. Instead, for $\lambda = 1$, phonon replicas spaced by $\omega$ are seen. A similar behavior occurs at $U = -1$: for small $\lambda$, $A(\epsilon)$ has two main peaks. Here, the electron-removal/addition parts of $A$ contribute to both peaks, since the e-ph interaction reorders the energy levels. 

The zero bias conductance $G$ is related to the spectral function~\cite{KurthJakob}. Using $v_{xc}$ from Eq.~(\ref{eq:vxc}), we calculate $G$ at zero temperature for a HH impurity connected to two 1D semi-infinite noninteracting leads. In this case, $G/G_0 = \sin^2(\pi n/2)$~\cite{Cornaglia}, with $n = (2/\pi)\arctan(-v_{KS}[n]/\gamma\pi) + 1$, $G_0$ the unit of quantum conductance, and $\gamma$ the level width in the wide-band limit. Results for $G$ as a function of $v$ and $\lambda$ are in 
Fig.~\ref{fig:single_site}(d),
where  $U = 1$ and $\mu = 0$. For $U' > 0$, $G$ has a plateau of width $U'$~\cite{GSSK}, but for $U' < 0$ we find a single narrow peak~\cite{Cornaglia,Perfetto12}. Overall, $G$ behaves smoothly as a function of the e-ph coupling, while the system evolves from the spin to the charge Kondo regime.


For further insight into e-e and e-ph interactions, in the SM 
we compare the exact double occupancy $\langle n_\uparrow n_\downarrow\rangle$ to the DFT one
(the latter is obtained via the single-site potential). While good agreement is found in many situations,  the comparison in the SM clearly suggests that kinetic-energy effects in $v_{xc}$ are important and thus 
a more general reference system than a single HH site should be used. This is taken into account in the next section.


{\it B) $v_{xc}$ from the infinite-dimensional Bethe lattice -}
\begin{figure}
 \includegraphics[width=0.95\columnwidth]{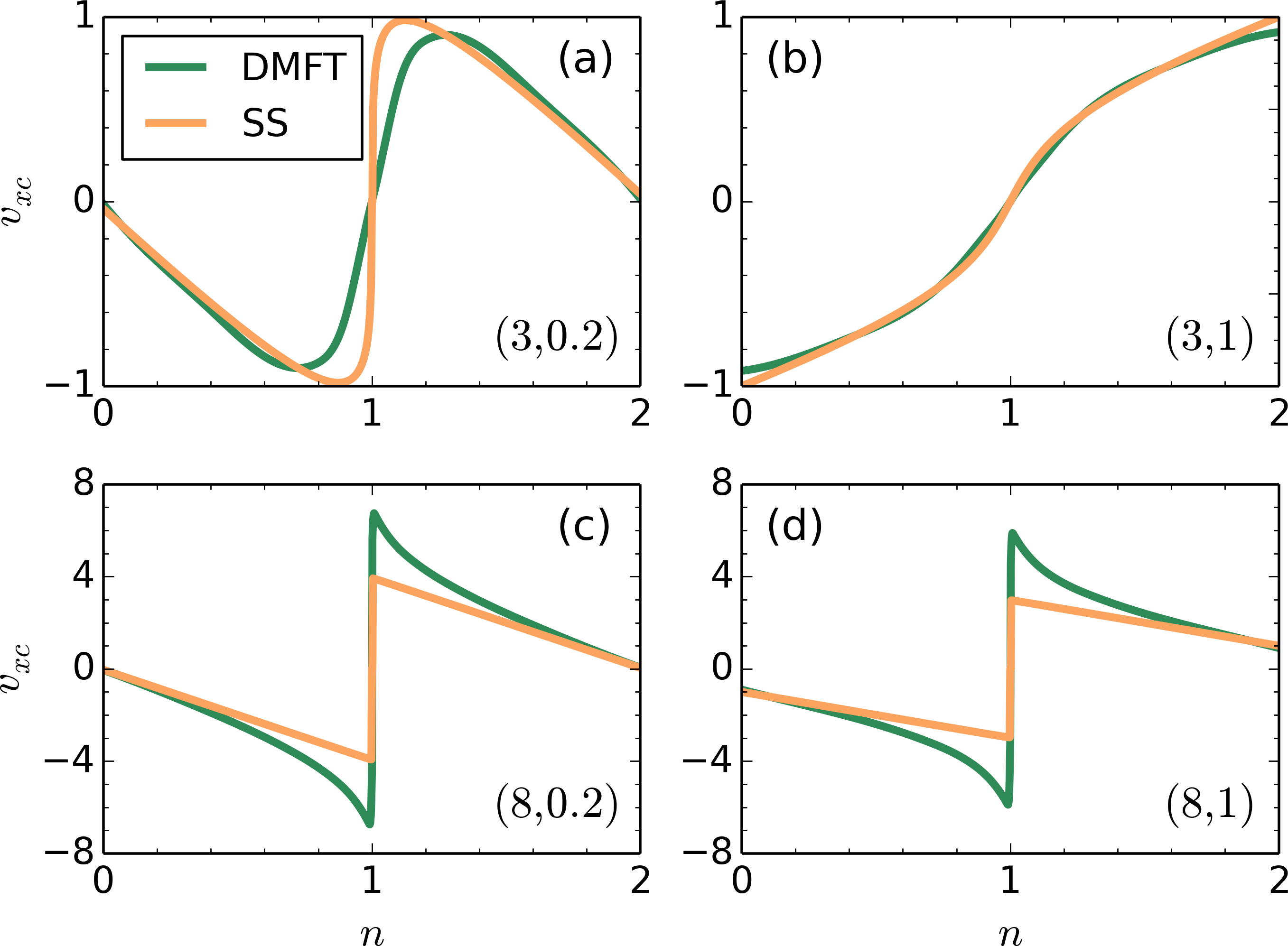}
 \caption{Exchange-correlation potential $v_{xc}$ as a function of electron density $n$, e-e interaction $U$ and e-ph interaction $g$. In panels (a)-(d), the  values of $(U,g)$ are shown for $\omega = 1$, $x = 0$, and $\eta_{Hxc} = g(n-1)$. Color coding: fitted DMFT data by polynomial interpolation (green curves) for $\beta = 200$; single Hubbard-Holstein site (SS) results (yellow curves). For the single-site potentials, $\beta$ is chosen via a fit to the DMFT results, giving $\beta \approx 5$ for $U = 3$ and $\beta\approx 100$ for $U = 8$.
}
 \label{fig:vxc}
\end{figure}
For the HH model on the $D=\infty$ Bethe lattice with bandwidth $4$ (in units of the hopping parameter), we estimate the ground state energy $E_{tot}$ within DMFT at $\beta = 200$~\cite{PhilippDMFT}. The exchange-correlation potential is explicitly determined for~ $U = 3, 8$ and $g = 0.2, 1$, corresponding in all cases to a screened interaction $U'>0$ \cite{future}.
After the calculation of $E_{xc}$ as a function of $n$ and $x$ via DMFT (see the SM), we compute the electron and phonon potential as $v_{xc}=\partial E_{xc}(n,x)/\partial n$ and $\eta_{xc} =  \partial E_{xc}(n,x)/\partial x$. To perform the necessary derivatives, we fit the DMFT data in $n \in [0,1]$ with piecewise fourth-order polynomials. For $n \in [1,2]$, we employ the symmetry  $E_{xc}(n) = E_{xc}(2-n)$. 

\begin{figure*}[tbph]
\includegraphics[width=0.245\textwidth]{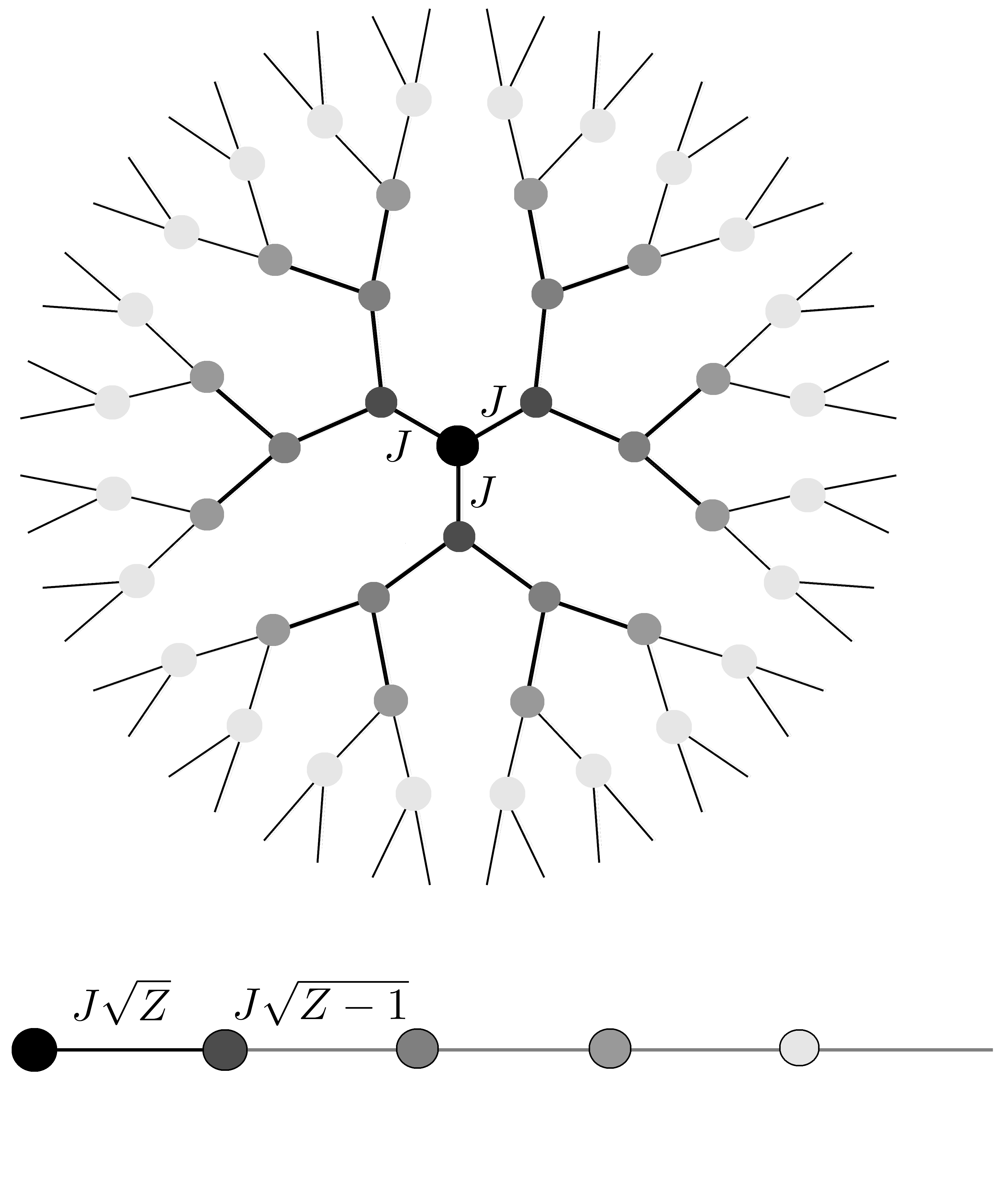}
\includegraphics[width=0.745\textwidth]{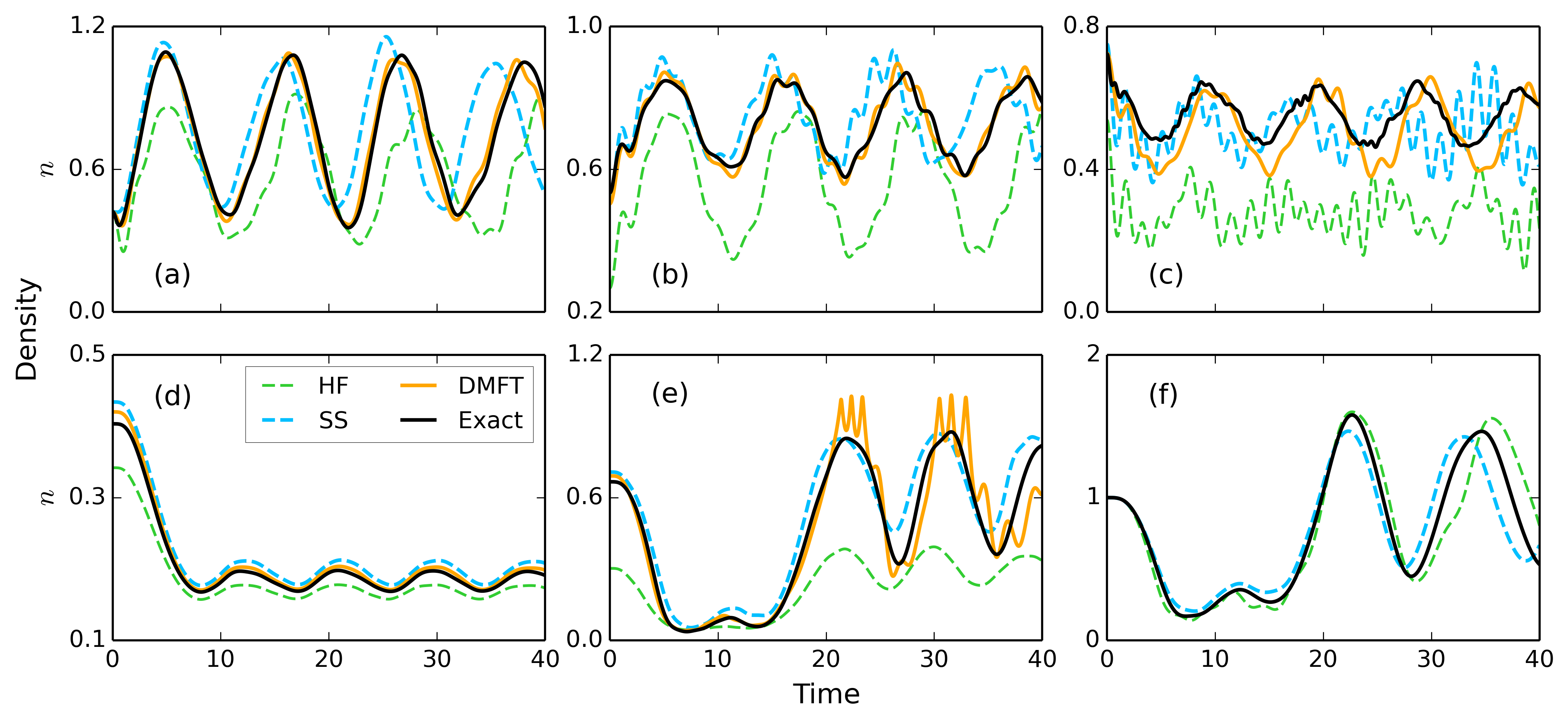}
 \caption{Left: Mapping of the Bethe lattice with coordination number $Z$ onto a one-dimensional linear chain. For finite $Z$, the effective hoppings in the chain are $J\sqrt{Z}$ between sites 0 and 1, and $J\sqrt{Z-1}$ otherwise. The results shown are for $Z \rightarrow \infty$, and renormalized hopping $J\rightarrow J/\sqrt{Z}$. Right: Dynamics of an 8 site Anderson-Holstein chain (see main text) with $n_\downarrow = n_\uparrow = 3$, for  $J = 1,~\omega = 1$, and $Z \rightarrow \infty$. In all panels, the optimization of $\beta$ gives a value $\beta \simeq 5$. Panels (a)-(c) correspond to a sudden quench of the interaction pair $(U, g)$; from left to right, $(0, 0) \rightarrow (3, 1), (8, 1) \rightarrow  (3, 1)$, and $(3, 1) \rightarrow  (8, 1)$, respectively. In panel (d), $U = 8$ and $g = 0.2$ and the external potential $v(t)$ is ramped to $v = 1$ in a time $T = 8$. For panels (e) and (f) the interactions are $U = 8$ and $g = 1$ or $U = 1$ and $g = 0.5$ respectively, and the external field is a pulse of strength $v = 1$ and duration $T = 4$.  For the explicit shape of  $v(t)$, see the SM.}  
 \label{Figure3}
\end{figure*} 

In Fig.~\ref{fig:vxc}(a)-(d) we show the $v_{xc}$ obtained from DMFT (for $E_{xc}$ results, see the SM). At $U=8$, $v_{xc}$ is discontinuous at $n=1$, but not at $U=3$. This is a DFT signature of the Mott-Hubbard transition in the HH model, in analogy with the purely electronic Hubbard model \cite{DKAPCV} (for the $D=\infty$ Bethe lattice, when $\beta \rightarrow \infty$, $U_{c1} \approx 4.7$, and $U_{c2} \approx 5.8$ \cite{Blumer}).   Interestingly, e-ph interactions not only renormalize the value of the XC discontinuity, but also ``delay" its onset. For further insight, in Fig.~\ref{fig:vxc} we also plot  by yellow lines the single-site results from Eq.~(\ref{eq:vxc}), with the value of $\beta$ fitted to best reproduce the DMFT curves. For panels (a) and (b), this gives $\beta = 5$, and a smeared XC discontinuity. Also, due to the small $U$ and large $g$ values, the shape of the single-site solution in panel (b) is dominated by the linear term $\frac{g^2}{\omega} \delta n$. In contrast, in panels (c) and (d) the fit gives $\beta=100$ for the single-site potential, already close to the zero temperature limit (where the discontinuity exists for all nonzero interactions $U'$). Overall, single-site and DMFT potentials agree for $n \in [0,1/2]$ but significant differences appear at higher fillings, with important consequences for time-dependent simulations.


{\it Bethe lattice mapping and real-time dynamics -} 
We now use the single-site and DMFT potentials for the real-time dynamics of an $L$-site chain with a HH impurity at one end (so-called Anderson-Holstein chain, see Fig.~\ref{Figure3}, left). This test system was chosen because the local density of states of a homogeneous Bethe lattice of coordination $Z$ and hopping term $J$ is identical to the one at the first site (site 0) of a semi-infinite chain. The mapping is obtained via Lanczos recursion (see SM), and also holds with HH interactions and time-varying fields at a single site of the Bethe lattice: in this case the local Green's function at that site is the same as the one at site 0 of the chain. When $Z\rightarrow \infty$, the chain Hamiltonian is
\begin{eqnarray}\label{hchain}
&H_\text{chain}& =  -J\sum_{i=0,\sigma}^\infty c_{i\sigma}^\dagger c_{i+1,\sigma} + \text{H.c.} \nonumber\\
&& +~v(t) \hat{n}_0+ U\hat{n}_{0\uparrow} \hat{n}_{0\downarrow}+\omega b^\dagger b +g \hat{n}_{0} (b^\dagger+b),
\end{eqnarray}
where $t$ labels time, $v(t)$ is a local perturbation, and the rescaling $J \rightarrow J/\sqrt{Z}$ keeps the hopping probability finite.

In the simulations, we use a finite chain of $L=8$ sites, which allows for exact numerical solutions. By virtue of the mapping, we are actually dealing with a $Z=\infty$ Bethe lattice truncated after eight layers and with one HH impurity in the center (Fig.~\ref{Figure3}). We consider $N_\uparrow = N_\downarrow= 3$ electrons in the chain (as before, $J = \omega= 1$). The system's time evolution is performed via exact diagonalization, as well as by TDDFT time propagation via the KS equations~\cite{TDDFTbook} within the adiabatic local density approximation (ALDA)~\cite{ALDA}. By setting $v_{xc}$ to zero, we also consider the Hartree-Fock (HF) dynamics.

Figure~\ref{Figure3}(a)-(c) shows the dynamics after a sudden interaction quench $(U_i, g_i)\rightarrow (U_f, g_f )$ at $t = 0$. This situation is within the scope of TDDFT, by freedom of choice of the initial state~\cite{RG84}. Further, quenches severely test the ALDA (typically employed within TDDFT, and used here). For the quench $(0, 0) \rightarrow (3, 1)$, panel (a) in Fig.~\ref{Figure3}, exact and TDDFT-DMFT results are in excellent agreement, while the single-site and HF solutions give a moderately good description. Instead, for  $(8, 1) \rightarrow (3, 1)$ and $(3, 1) \rightarrow (8, 1)$, panels (b) and (c), the agreement worsens, due to stronger interactions. However, the DMFT potential still qualitatively performs well, while the HF solution fails to capture the main features. 

Figure~\ref{Figure3}(d)-(f) shows the dynamics induced by an external field $v(t)$. In panel (d), where $U = 8$ and $g = 0.2$, $v(t)$ is ramped on in a time $T = 8$, and kept constant (=1) afterwards (see the SM). There is excellent agreement between exact and TDDFT results. In (e, f) $v(t)$ is a soft square pulse of duration $T = 4$ and amplitude $v = 1$, switched on and off in a time $T = 8$. In (e), where $U = 8$ and $g = 1$, the DMFT potential initially gives a very good agreement but this worsens near $n=1$. This is a known behavior, due to the discontinuity of $v_{xc}$ at $n = 1$~\cite{Kurth10,DKAPCV}. Finally, we briefly turn to the $U'<0$ region, using the single-site potential (the case of the DMFT potential is left for future work). The results of panel (f), where $U =-1$ and $g = 0.5$, suggest that (TD)DFT can also be used for the attractive regime, but better potentials than the single-site one are clearly needed. 

{\it Conclusions.- } By means of a two-component density functional theory (DFT), we have introduced a novel approach to the Hubbard-Holstein (HH) model, a popular template to study e-e and e-ph interactions in lattice systems. We also explicitly determined and characterized electron and phonon exchange-correlation (XC) potentials, analytically for a simple one-site system, and via dynamical mean field theory for a Bethe lattice in $D=\infty$. Comparisons between DFT and exact results showed that the newly found potentials perform very well across an appreciable range of interaction strengths and electron densities. 

Possible directions for immediate extensions include the analysis of the phonon overscreening regime, and a formulation for lattices in $D\leq3$ or for the linear response regime. Application-wise, an appealing option would be to explore how phonon-like degrees of freedom affect the physics of cold atoms in optical lattices (e.g. cloud expansion after trap removal, disorder effects, entanglement distillation, etc.). Finally, a key development would be the introduction of memory and non local effects in the XC potentials, by exploiting connections to many-body approximations within Green's function schemes.

\begin{acknowledgments}
We thank C.-O Almbladh and A. Privitera for discussions. 
E. V. B and P. H. were supported from Crafoordska stiftelsen.
 P. W. acknowledges support from ERC Consolidator Grant No. 724103 
and SNSF Grant No. 200021-165539.
C. V. was supported by the Swedish Research Council.
\end{acknowledgments}


\begin{thebibliography}{999}
\bibitem{future} A systematic investigation of the $U'<0$ regime is left to future work.
\end{thebibliography}
\end{document}